\documentclass[aps,prl,twocolumn,preprintnumbers,superscriptaddress,10pt,longbibliography]{revtex4-2}

\usepackage[T1]{fontenc}
\usepackage[utf8]{inputenc}

\usepackage{amssymb}
\usepackage{amsmath}
\usepackage{amsfonts}
\usepackage{graphicx}
\usepackage{subfigure}
\usepackage{color}
\usepackage{dcolumn}
\usepackage{hyperref}

\newcommand{\eq}{\begin{equation}}
\newcommand{\eqx}{\end{equation}}
\newcommand{\eqn}{\begin{eqnarray}}
\newcommand{\eqnx}{\end{eqnarray}}
\newcommand{\be}{\begin{equation}}
\newcommand{\ee}{\end{equation}}
\newcommand{\bea}{\begin{eqnarray}}
\newcommand{\eea}{\end{eqnarray}}

\newcommand{\f}[2]{\frac{#1}{#2}}

\newcommand{\LL}{{\mathcal L}}

\newcommand{\eps}{\varepsilon}
\newcommand{\al}{\alpha}

\newcommand{\Sg}{\Sigma}
\newcommand{\dl}{\delta}
\newcommand{\Dl}{\Delta}

\newcommand{\gm}{\gamma}
\newcommand{\Gm}{\Gamma}

\newcommand{\qqqq}{\quad\quad\quad\quad}

\begin{document}

\title{A perfect fluid hydrodynamic picture of domain wall velocities at strong coupling}

\author{Romuald A. Janik}
\email{romuald.janik@gmail.com}
\affiliation{Institute of Theoretical Physics, Jagiellonian University, Lojasiewicza 11, 30-348  Krakow, Poland}
\author{Matti J\"arvinen}
\email{matti.jarvinen@apctp.org}
\affiliation{Asia Pacific Center for Theoretical Physics, Pohang 37673, Republic of Korea}

\affiliation{Department of Physics, Pohang University of Science and Technology, Pohang 37673, Republic of Korea}
\author{Hesam Soltanpanahi}
\email{hesam.soltanpanahi@uj.edu.pl}
\affiliation{Institute of Theoretical Physics, Jagiellonian University, Lojasiewicza 11, 30-348  Krakow, Poland}

\author{Jacob Sonnenschein}
\email{cobi@tauex.tau.ac.il}
\affiliation{The Raymond and Beverly Sackler School of Physics and Astronomy, Tel~Aviv University, Ramat Aviv 69978, Israel}
\affiliation{Simons Center for Geometry and Physics, SUNY, Stony Brook, NY 11794, USA}

\begin{abstract}
We show that for a range of strongly coupled theories with a first order phase transition, the domain wall or bubble velocity can be expressed in a simple way in terms of a perfect fluid hydrodynamic formula, and thus in terms of the equation of state. We test the predictions for the domain wall velocities using the gauge/gravity duality.
\end{abstract}

\preprint{APCTP Pre2022 - 008}

\maketitle


\noindent {\bf Introduction.}
The study of the dynamics of the expansion of nucleated bubbles in a theory with a $1^{st}$ order phase transition has recently gained renewed interest in view of gravitational wave detection experiments \cite{Demidov:2017lzf, Chala:2018ari, Ahriche:2018rao, Caprini:2019egz, Caldwell:2022qsj}, as the collisions and coalescence of bubbles as well as the accompanying plasma in the early universe may act as a source of potentially detectable gravitational waves.

In this context, a key parameter of interest is the velocity of the bubble wall. Despite the fact that there is a pressure difference between the two phases, any initial accelerated motion very soon stabilizes to a motion with a uniform velocity. The conventional explanation is that the net force is balanced by the friction of the metastable phase, which is, however, very challenging to compute (see~\cite{Moore:1995ua, Espinosa:2010hh,Hindmarsh:2020hop} as well as~\cite{DeCurtis:2022hlx,Laurent:2022jrs} and references therein for recent work in this direction). From this perspective, the domain wall velocity is thus a consequence of nonequilibrium dynamics of the theory.

In this letter we argue that for a range of strongly coupled theories which have a holographic dual, the physics of the uniform motion of the domain wall is much simpler and does not require discussing the nontrivial nonequilibrium or dissipative regime of the theory. Indeed, we show that the domain wall velocity may be reliably computed using just perfect fluid hydrodynamics and thus the equation of state.
We test the predictions using holographic modelling of the process of bubble expansion both for nucleated bubbles of a stable phase within an overcooled medium and for an interface between two phases at different temperatures. 
In order to be able to compare with numerical holographic simulations, all the domain walls and bubbles that we consider have a planar symmetry.
The theories we model are a confining/deconfined system and a bottom-up holographic model with two deconfined phases. We also compare our predictions with some holographic results of~\cite{Bea:2021zsu}.


\bigskip

\noindent {\bf The holographic frameworks.}
A  method of describing confinement in holographic settings is to start from a geometry dual to a 4+1 dimensional QFT and compactify one of the spatial dimensions on a circle~\cite{Itzhaki:1998dd,Brandhuber:1998er,Witten:1998zw}. Then the confining geometry takes the form of a cigar in the compactified and holographic directions. The circumference  of the cigar is identified as the scale of confinement in the resulting 3+1 dimensional field theory.
This geometry is usually called the Witten model in the literature. 
In the case of 2+1 dimensional field theory, 
the analogous confined geometry is the AdS$_5$ soliton~\cite{Horowitz:1998ha}. The deconfinement transition takes place between the soliton and the thermal AdS$_5$ geometry, i.e., a planar black hole (with one compactified spatial coordinate)~\cite{Aharony:2006da}. Constructing the domain wall between the soliton and the black hole is challenging because the topologies of the two geometries differ, but has been found numerically by Aharony, Minwalla and Wiseman (AMW) \cite{Aharony:2005bm}.
An analogous solution for the original Witten model has not been constructed so far, hence we focus here on the lower dimensional version (which we will anyhow refer to as the Witten model in the following).

In \cite{Janik:2021jbq}, we found that the energy-momentum tensor corresponding to the AMW domain wall solution can be 
accurately described using a boundary field theory description involving hydrodynamic degrees of freedom as well as an additional scalar field $\gm$ which describes the interpolation between the two phases with $\gm=0$ corresponding to the deconfined phase and $\gm=1$ to the confined one. The Lagrangian of the model is
\eqn
\label{e.simp1}
\LL &=& (1-\Gm(\gm))p(T) + \Gm(\gm) \\ \nonumber &&-\f{1}{2} c(T,\gm) (\partial \gm)^2 - d(T,\gm) \gm^2 (1-\gm^2)
\eqnx
with $\Gm(\gm) = \gm^2 (3-2\gm)$.
The energy-momentum tensor is given by
\eq
\label{e.simp3}
T_{\mu\nu} = (1-\Gm(\gm))\, T_{\mu\nu}^\mathrm{confining} + \Gm(\gm)\, T_{\mu\nu}^\mathrm{deconfined} + T_{\mu\nu}^\Sg
\eqx
Explicit formulas for the energy-momentum tensors and the coefficients in (\ref{e.simp1}) are given in the \textit{Supplemental material}. 
See also~\cite{Bigazzi:2020phm,Ares:2021ntv,Ares:2021nap} for  different effective descriptions for first order phase transitions in holography. 
The importance of the above simplified description lies in the fact that a direct time-dependent numerical relativity simulation of the Witten model is extremely difficult due to the different topology of the two phases~\cite{Bantilan:2020pay}. Hence in the Witten model we will perform numerical evolution using (\ref{e.simp1})-(\ref{e.simp3}).

\begin{figure}[t]
	\begin{center}
	\includegraphics[width=.45\textwidth]{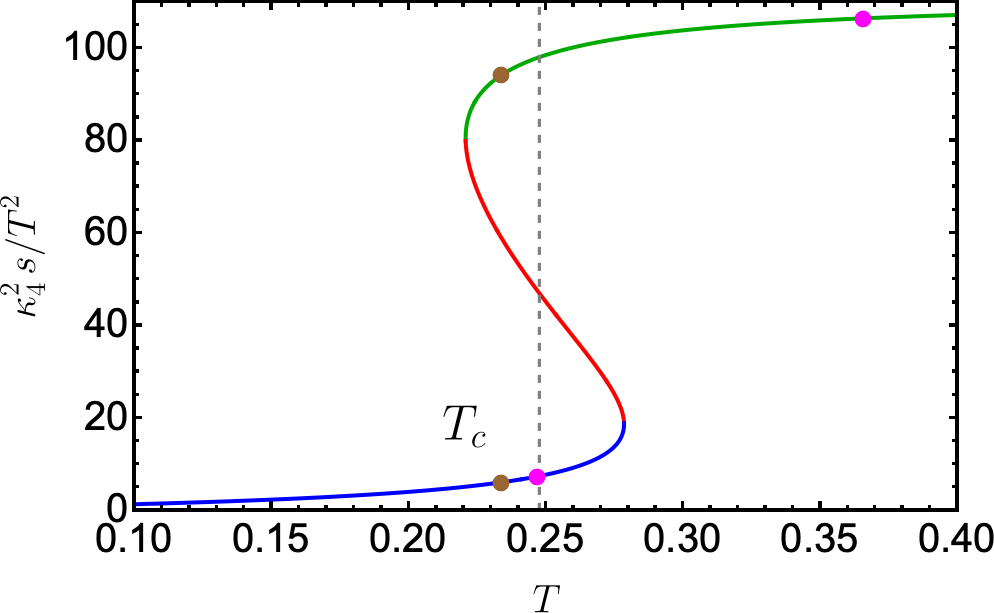}
	\caption{Equation of state of the nonconformal model. Black dots indicate sample phases in the nucleated bubble simulations, red dots indicate sample phases for the expanding interface between phases at different temperatures.}
	\label{fig.eos}
	\end{center}
\end{figure}

As a complementary system we consider a bottom-up holographic gravity+scalar model in 4D bulk 
\eq
\label{e.nonconformal}
S=\frac{1}{2\kappa_4^2}\int d^4x \sqrt{-g}  \left[ R-\frac{1}{2}\, \left( \partial \phi \right)^2 - V(\phi) \, \right]
\eqx
with the potential
\eq
\label{e.potential}
V(\Phi) = -6 \cosh\left(\f{\Phi}{\sqrt{3}}\right) - 0.2\, \Phi^4
\eqx
exhibiting a $1^\textrm{st}$ order phase transition which was studied in \cite{Janik:2017ykj, Bellantuono:2019wbn}. In this model both phases are deconfined and have a holographic black hole description.
The equation of state is shown in Fig.~\ref{fig.eos}.
In this case we can perform full-fledged numerical relativity simulations which will allow us to confirm some qualitative features of the domain wall expansion deduced from the simplified model simulations. Since for this theory we can study the full holographic evolution, our treatment retains all possible non-equilibrium and dissipative features in the dynamics. We will refer to this model as the non-conformal holographic model.


\begin{figure}[h]
	\begin{center}
	\includegraphics[width=0.23\textwidth]{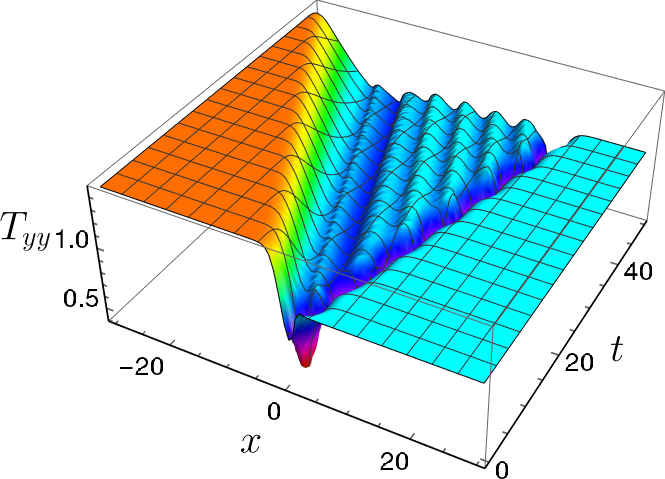}
	\includegraphics[width=0.23\textwidth]{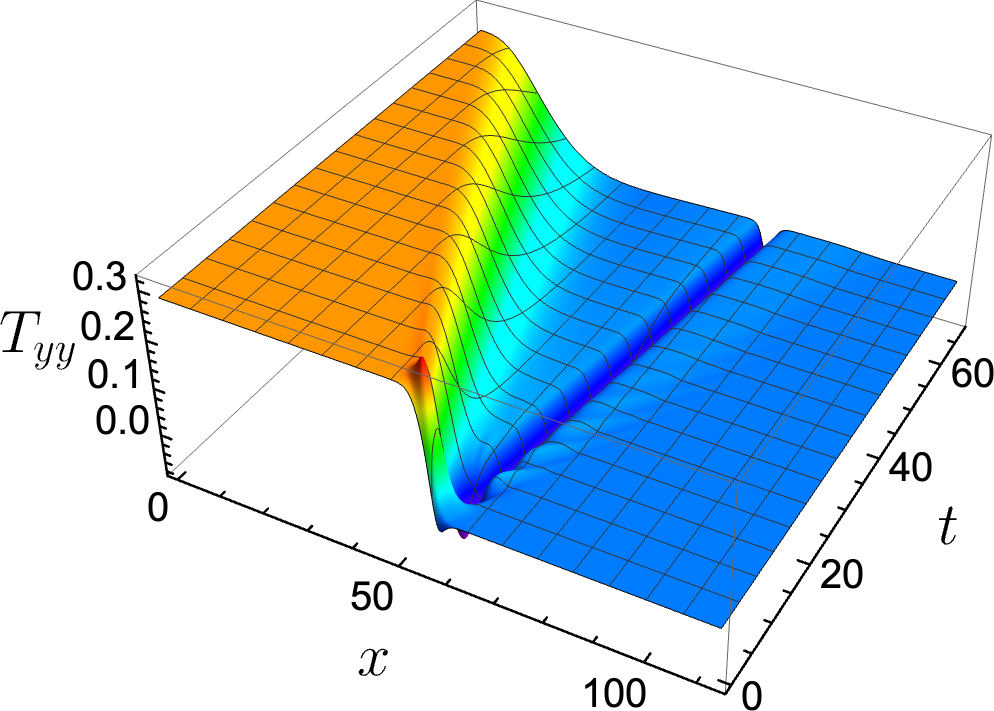}
	\includegraphics[width=0.23\textwidth]{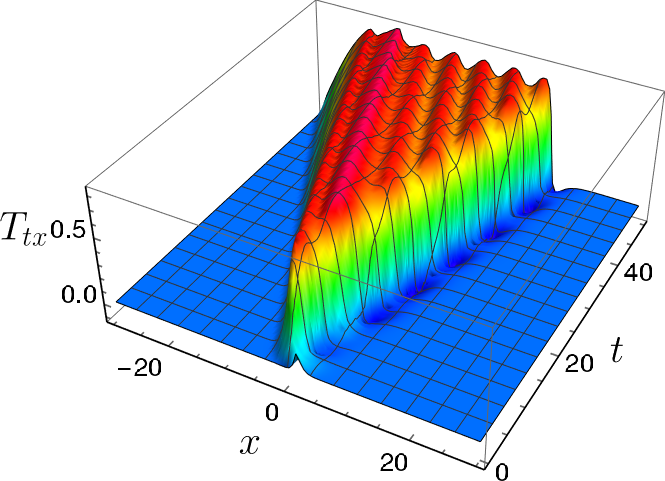}
	\includegraphics[width=0.23\textwidth]{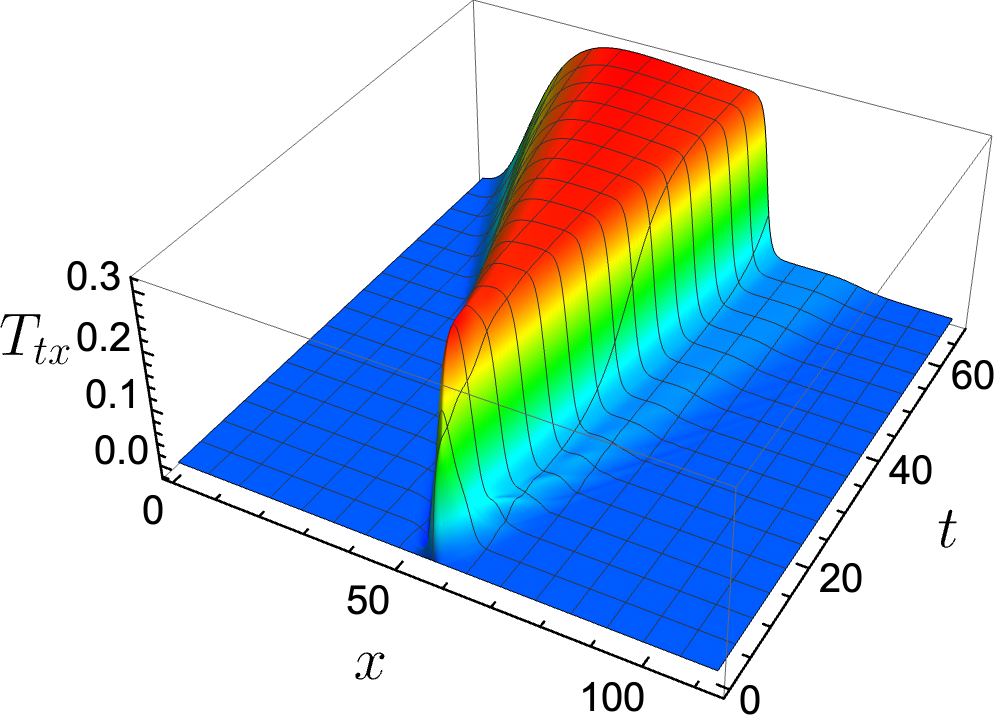}
\caption{Evolution of pressure $T_{yy}$ (top row) and the momentum flow $T_{tx}$ (bottom row) in the Witten model (left) and nonconformal model (right). The high entropy (deconfined) phase is on the left with the domain wall moving to the right (clearly seen as a dip of $T_{yy}$).}
	\label{fig.Tyy}
	\end{center}
\end{figure}

\bigskip
\noindent {\bf Qualitative structure of domain wall motion at strong coupling.} Let us first study the evolution of the interface between two phase domains: deconfined phase at $T>T_c$ and the confined phase. The pressure of the relevant phases can be directly read off from the $T_{yy}$ component of the energy-momentum tensor (where $y$ is the coordinate along the domain wall). In Fig.~\ref{fig.Tyy}(top left) we show the time evolution of  $T_{yy}$ in the simplified model (\ref{e.simp1})-(\ref{e.simp3}).

Since the confined phase in the Witten model does not depend on the temperature, one could interpret this configuration as either a system at $T>T_c$ or as an interface between a high entropy phase at $T>T_c$ and the low entropy phase at $T=T_c$. In Fig.~\ref{fig.Tyy}(top right) we show a plot of a counterpart of the latter configuration in the nonconformal holographic model.

The conventional picture explaining a constant domain wall velocity despite the imbalance of pressures on both sides of the domain wall is that the net force coming from the difference of pressures is balanced by friction induced by the phase in front of the moving domain wall \cite{Hindmarsh:2020hop}. This makes the determination of the domain wall velocity a very challenging problem.

The holographic simulations shown in Fig.~\ref{fig.Tyy} show, however, a quite different picture which arises at strong coupling. Firstly, the pressures on both sides of the domain wall are in fact very close to each other. Therefore, the domain wall motion with constant velocity is in fact very natural here and does not need any balancing friction force.
Secondly, the pressure difference between the phases is supported on a hydrodynamic wave moving in the high entropy phase \emph{away} from the domain wall~\footnote{ 
The models differ in the intermediate region where the Witten's model shows wavy behavior. Apparently the waves 
are absent in the nonconformal model because it includes dissipation, whereas for the Witten's model we use a perfect fluid description.}.

Plotting the $T^{tx}$ component of the energy-momentum tensor in the bottom of Fig.~\ref{fig.Tyy}, we see that the high entropy phase to the left of the domain wall is not static and has a flow velocity in the same direction as the motion of the domain wall. In fact, as we will argue shortly, for the studied holographic systems the flow velocity is very close to the domain wall velocity.

\bigskip
\noindent {\bf Perfect fluid description.} From the numerical simulations discussed above, we observe that the late time configuration consists of a domain wall moving with a constant velocity with balanced pressures on both sides and a hydrodynamic wave moving in the opposite direction supporting the pressure difference. We can therefore try to glue together the two ingredients.

In the Witten model, there exists an exact holographic solution for a moving domain wall, which is the static AMW domain wall geometry boosted to the domain wall velocity. Since in the AMW domain wall solution, the fluid in the deconfined phase is at rest, this implies that in the boosted solution, the velocity of the fluid would be equal to the velocity of the domain wall. This is a crucial relation as it translates the difficult problem of computing the domain wall velocity into computing the fluid velocity in the high entropy deconfined phase.

In addition, the pressure of the boosted solution would be equal to the pressure at the $1^{st}$ order phase transition $p=p_c \equiv p(T_c)$.

For the consistency of this picture it is important to check that
the (boosted) AMW domain wall is the only static (moving) domain wall solution in our hydrodynamic description of the Witten model. 
We studied numerically in the rest frame of the wall such solutions of (\ref{e.simp1}), where the fluid has nonzero velocity on at least one side of the wall, and only found interpolating solutions when the velocity vanishes on both sides, i.e., the standard domain wall solution in the rest frame.

So we are left with finding a hydrodynamic solution interpolating between a static plasma with $p=p_A>p_c$ and a moving plasma with $p=p_c$ and velocity $v$. We would like to express $v$ as a function of the pressure difference.
It is illuminating to first consider the problem in the linearized approximation around some reference point:
\eq
p = p_\mathrm{ref} + \dl p \qqqq u^\mu=(\cosh \al, \sinh\al, 0)
\eqx
where both $\delta p$ and $\al$ are small. 
Then the solution for the perfect fluid hydrodynamics for a left-moving wave is
\eq
\dl p = f(x+ c_s t) \qqqq \al = -\f{f(x+c_s t)}{(\eps_\mathrm{ref} +p_\mathrm{ref}) c_s} + \mathrm{const}
\eqx
Implementing the above boundary conditions for the pressures and velocities we obtain
\eq
\label{e.linearized}
v_\mathrm{domain\ wall} = v_\mathrm{fluid} = \tanh  \f{\Dl p}{(\eps_\mathrm{ref} +p_\mathrm{ref}) c_s}
\eqx
It turns out, however, that nonlinear hydrodynamic effects are important. The generalization is called \emph{a simple wave} \cite{landau2013fluid} (see also \textit{Supplemental material}) and the counterpart of (\ref{e.linearized}) is
\eq
\label{e.nonlinear}
v_\mathrm{domain\ wall} = \tanh \int_{p_{c}}^{p_A} \f{dp}{(\eps + p)c_s}  \ \equiv\ \tanh \int_{T_{c}}^{T_A} \f{dT}{T c_s} 
\eqx
Note that the formula is expressed completely in terms of equation of state data.

We have checked numerically that both the velocity of the moving plasma and the velocity of the domain wall in the Witten model agree well with~\eqref{e.nonlinear} at small pressure differences: 
\eq
v_\mathrm{domain\ wall} \approx v_\mathrm{fluid} \approx \sqrt{3}\, \frac{\Delta T}{T}
\eqx
within the precision of less than half a percent.

\begin{figure}[t]
	\begin{center}
	\includegraphics[width=.47\textwidth]{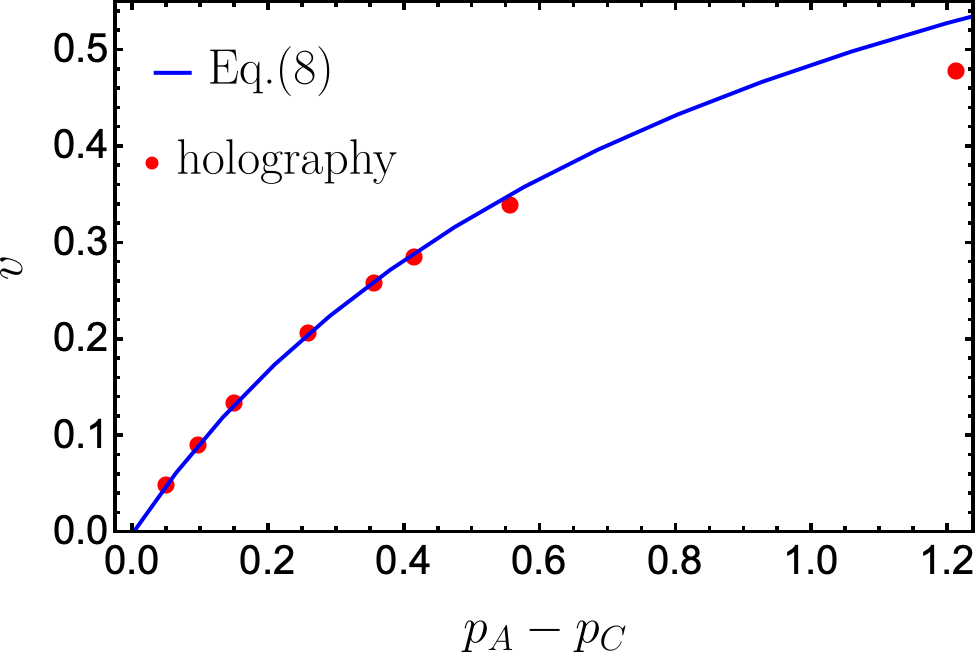}
	\caption{Domain wall velocities obtained from holographic simulations in the nonconformal model as a function of pressure difference compared with formula (\ref{e.nonlinear}) evaluated in the high entropy phase. The high entropy phase is at a temperature $T_A>T_c$. The low entropy phase is at $T_c$.}
	\label{fig.highenergy}
	\end{center}
\end{figure}

An analogous discussion for the nonconformal model with two deconfined phases leads to a couple of subtleties. In this case, both phases of the theory have a hydrodynamic description, and we are faced with a choice which phase (if any) to use for the hydrodynamic formula (\ref{e.nonlinear}). Moreover, an examination of our numerical holographic simulations indicate that the stationary moving domain wall will no longer be just a boosted version of the static solution as fluid velocities on both sides of the domain wall are not equal.

To understand this issue, it is convenient to pass to the domain wall rest frame and use the standard relations following from energy-momentum conservation which link the hydrodynamic parameters of the two fluids on both sides of the domain wall \cite{Gyulassy:1983rq, Espinosa:2010hh}:
\eq
\label{e.esposito}
\f{v_H}{v_L} = \f{\eps_L + p_H}{\eps_H + p_L}  \qqqq v_H v_L = \f{p_H - p_L}{\eps_H - \eps_L}
\eqx
where the subscripts $H$ and $L$ denote the high and low entropy phase respectively. Let us evaluate $v_H$ from the first equation
\eq
v_H = \f{\eps_L + p_H}{\eps_H + p_L} v_L < \f{\eps_L + p_H}{\eps_H + p_L}
\eqx
where the inequality follows from $v_{L,H}<1$. The last ratio can be evaluated from the equation of state. Assuming that the pressures $p_L \sim p_H$ (as seen in Fig.~\ref{fig.Tyy}(top right)), the ratio is approximately equal to the ratio of entropies of the two phases (using $Ts=\eps+p$). In the case of our model this is a small number, similarly for any confinement/deconfinement system due to the scaling with $N_c$. Since $v_H$ is small in the rest frame of the domain wall, this means that in the laboratory frame, the fluid velocity of the high entropy phase should be close to the domain wall velocity. Hence the conclusion is that the formula (\ref{e.nonlinear}) should be evaluated in the high entropy phase of the system.

In Fig.~\ref{fig.highenergy}, we compare the formula (\ref{e.nonlinear}) to the domain wall velocities obtained from the holographic simulations of the nonconformal model (\ref{e.nonconformal}). We find a very good agreement for a significant range of pressure differences with deviations occurring only for a quite large velocity. 

\begin{figure}[t]
	\begin{center}
	\includegraphics[width=.4\textwidth]{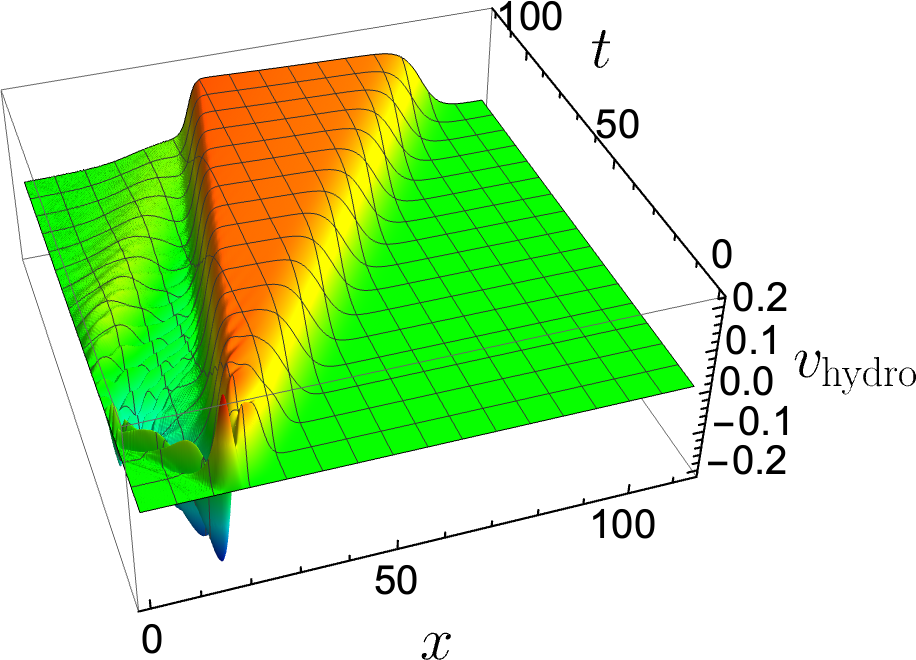}
	\caption{Perfect fluid velocities extracted from the energy momentum tensor for the evolution of a nucleated bubble in an overcooled medium.}
	\label{fig.nucleated}
	\end{center}
\end{figure}

\bigskip
\noindent {\bf Nucleated bubbles.} Up until now we have discussed the motion of a domain wall separating two (potentially infinite) phases at different temperatures. We will now move on to the physically more relevant case of the expansion of a nucleated bubble. As the initial conditions we take a small bubble of the stable low entropy phase in an overcooled medium of the high entropy phase at the same temperature (c.f. Fig.~\ref{fig.eos}).  

A sample velocity profile extracted from the holographic simulation is shown in Fig.~\ref{fig.nucleated}. Since the high entropy overcooled phase has smaller pressure than the stable phase, the hydrodynamic simple wave is now in front of the domain wall and moving in the same direction. This is in contrast to the case of the domain wall between high entropy phase with $T>T_c$ and the low entropy phase with $T=T_c$, which we discussed above.

\begin{figure}[t]
	\begin{center}
	\includegraphics[width=.45\textwidth]{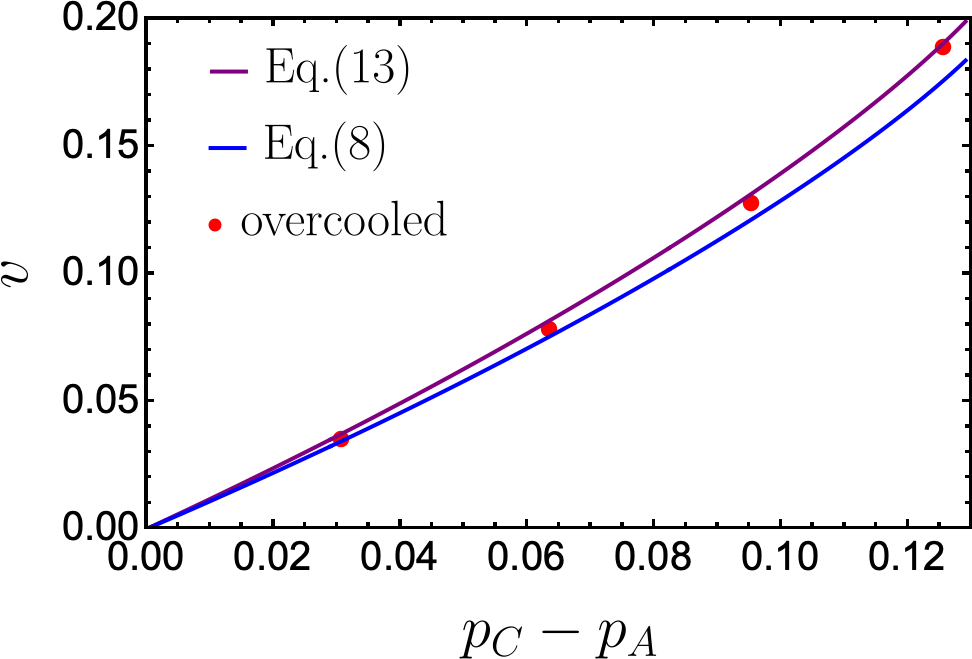}
	\caption{Nucleated bubble domain wall velocities obtained from holographic simulations in the nonconformal model as a function of pressure difference compared with formula (\ref{e.corrected}) evaluated in the high entropy phase.
	}
	\label{fig.bubbles}
	\end{center}
\end{figure}

As the velocity profile extracted from the holographic simulations indicates that the phase inside the bubble is at rest (this is the so-called deflagration case), we can improve in this case our formula (\ref{e.nonlinear}). Indeed using the notation from (\ref{e.esposito}), we see that $v_L=-v_\mathrm{domain\; wall}$, hence computing $v_H$ from (\ref{e.esposito}) and passing back to the laboratory frame we have
\eq
v_\mathrm{domain\; wall} = \f{1}{ 1- \f{\eps_L + p_H}{\eps_H + p_L} } v_\mathrm{fluid} 
\eqx
where $v_\mathrm{fluid}$ is given by the hydrodynamic formula in the high entropy phase. Hence we set
\eq
\label{e.corrected}
v_\mathrm{domain\; wall} = \f{1}{ 1- \f{\eps_L + p_H}{\eps_H + p_L} } 
\tanh \int_{p_{A}}^{p_C} \f{dp}{(\eps + p)c_s}
\eqx
where $p_C$ is the pressure inside the nucleated bubble, while $p_A$ is the pressure of the overcooled environment.
The formula (\ref{e.corrected}) is compared with numerical results from holography in Fig.~\ref{fig.bubbles}.

\begin{figure}[t]
	\begin{center}
	\includegraphics[width=0.23\textwidth]{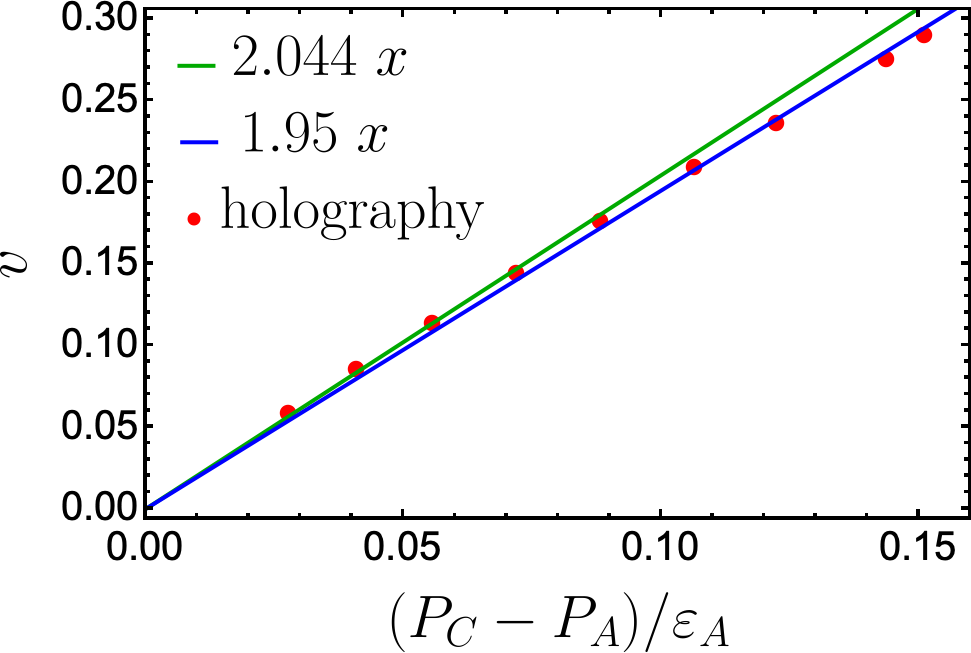}
	\includegraphics[width=0.23\textwidth]{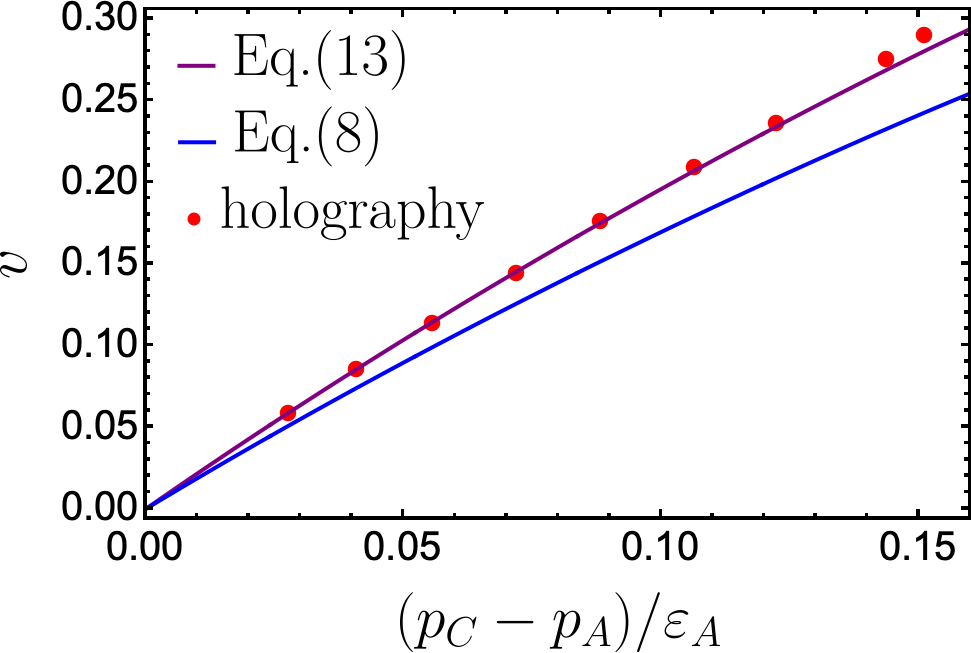}
	\caption{Data from Fig. 7 of \cite{Bea:2021zsu} compared with the linear formula (\ref{e.linearcoef} and the original fit from \cite{Bea:2021zsu} (left), and with our formula (\ref{e.corrected}).}
	\label{fig.fig7}
	\end{center}
\end{figure}

An interesting phenomenological formula for the nucleated bubble domain wall velocity in a holographic theory was proposed in \cite{Bea:2021zsu}:
\eq
v = \text{const}\cdot \f{\Dl p}{\eps_A}
\eqx
where the proportionality constant was fitted to be equal to 1.95 for a bottom-up scalar+gravity model (with 5D bulk) -- see Fig. 7 in \cite{Bea:2021zsu}. Our formula (\ref{e.corrected}) does not lead to an exactly linear dependence, but we can evaluate the slope coefficient. Indeed, for small pressure differences we are close to $T_c$, and hence we can evaluate (\ref{e.corrected}) with $p_L=p_H=p_c$. Keeping in mind that in this limit $\eps_H \simeq \eps_A$ we get
\eq
\label{e.linearcoef}
v_\mathrm{domain\; wall}^\mathrm{linearized} = \underbrace{\f{\eps_H}{\eps_H - \eps_L} \f{1}{c_s}_{| T=T_c}}_\text{const} \cdot \f{\Dl p}{\eps_A}
\eqx
For the theory considered in \cite{Bea:2021zsu}, the constant can be evaluated to be $2.044$. In Fig.~\ref{fig.fig7}(left) we compare the two linear coefficients to the data from \cite{Bea:2021zsu}. We see that for small velocities, the larger coefficient seems to work better.  
In Fig.~\ref{fig.fig7}(right), we compare the same data with our full nonlinear formula (\ref{e.corrected}).

\begin{figure}[t]
	\begin{center}
	\includegraphics[width=.45\textwidth]{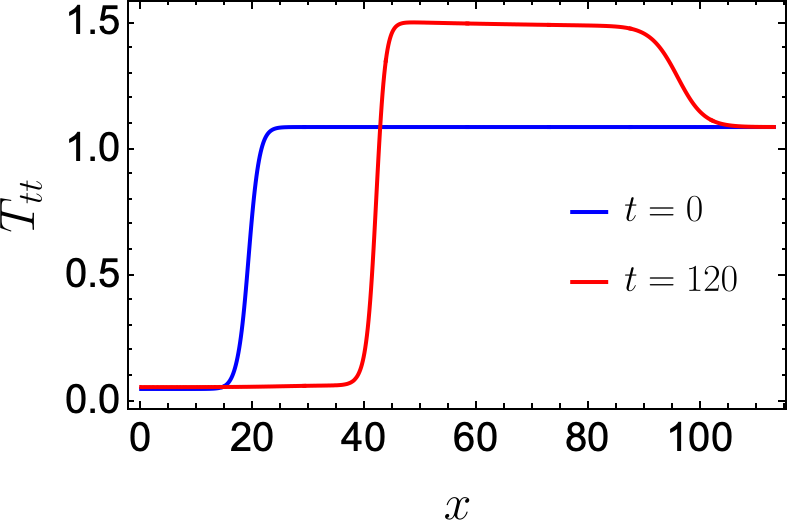}
	\caption{Energy density of the initial conditions for a nucleated bubble, and after a period of evolution. The latent heat is transformed into the energy of the perfect fluid hydrodynamic wave.}
	\label{fig.energybalance}
	\end{center}
\end{figure}

An interesting physical aspect of the nucleated bubble expansion is the energy balance and the fate of the latent heat which gets released when the bubble of the stable vacuum expands. 
We find that it gets transformed into the perfect fluid wave~\footnote{The pressure of the perfect fluid wave is close to the pressure inside the bubble, hence its temperature is typically higher and closer to $T_c$ than the temperature of the asymptotic overcooled plasma.} moving in front of the domain wall. This is illustrated in Fig.~\ref{fig.energybalance}. 

\bigskip
\noindent {\bf Discussion.}
In this paper we showed that for various strongly coupled theories with a holographic dual, the domain wall velocities may be expressed just in terms of the properties of the equation of state, employing perfect fluid hydrodynamics.
The resulting predictions agree quite well with direct holographic simulations of a moving interface between $T_A>T_c$ and $T_c$ (see Fig.~\ref{fig.highenergy}) and nucleated bubbles (see Fig.~\ref{fig.bubbles} and \ref{fig.fig7}).

\textit{A-posteriori}, since the final formulas (\ref{e.nonlinear}) and (\ref{e.corrected}) are expressed in terms of the equation of state, holography is not really needed for the evaluation of the domain wall velocity. For us holography was, however, crucial in order to check the validity of our formulas.
We expect that the key requirements for the applicability of our predictions is the existence of a hydrodynamic description with a low viscosity as well as a relatively large ratio of entropies of the two phases.

This work leads to many open questions related to  the velocities and other  properties of  domain walls: i) Our results relate to  a variety of physical systems including confinement and deconfinement, deconfined phase  and an overcooled one, conformal and nonconformal theories, related to holographic top-down and bottom-up scenarios and  to gauge theories   in 2+1 and 3+1 dimensions. To what extent does this simple dynamics of the domain wall apply to  other systems of first order phase transition? ii) We expect that it should be possible to generalize 
the results to systems with circular/spherical bubbles (c.f. holographic simulations in \cite{Bea:2022mfb}). iii) We addressed pure gauge theories. Incorporating flavored quark degrees of freedom, which in the holographic picture means adding probe flavor branes, is of utmost importance. Some interesting work in this direction has been recently done in~\cite{Bigazzi:2020phm,Bigazzi:2021ucw}.  iv)
The transition from a  deconfined  to a confined phase is achieved  via hadronization, whose holographic description remains elusive.  It is not clear how this  affects the dynamics of the domain wall. v) There might be certain  possible applications of our results  to heavy ion collisions and quark gluon plasma, potentially at non-zero density.
It would be also very interesting to investigate possible applications for the early universe.


\bigskip
\noindent{\bf Acknowledgements.} 
We would like to thank David Mateos for his comments and for providing us with interesting references and Larry McLerran, Govert Nijs for discussions. RJ would like to thank the Simons Center for Geometry and Physics for hospitality during the final stage of this work. MJ has been supported 
by an appointment to the JRG Program at the APCTP through the Science and Technology Promotion Fund and Lottery Fund of the Korean Government. MJ has also been supported by the Korean Local Governments -- Gyeong\-sang\-buk-do Province and Pohang City -- and by the National Research Foundation of Korea (NRF) funded by the Korean government (MSIT) (grant number 2021R1A2C1010834). The work of JS was supported  by a Center  of excellence of the Israel Science Foundation (grant number
2289/18)


\bibliography{references}

\begin{thebibliography}{30}%
\makeatletter
\providecommand \@ifxundefined [1]{%
 \@ifx{#1\undefined}
}%
\providecommand \@ifnum [1]{%
 \ifnum #1\expandafter \@firstoftwo
 \else \expandafter \@secondoftwo
 \fi
}%
\providecommand \@ifx [1]{%
 \ifx #1\expandafter \@firstoftwo
 \else \expandafter \@secondoftwo
 \fi
}%
\providecommand \natexlab [1]{#1}%
\providecommand \enquote  [1]{``#1''}%
\providecommand \bibnamefont  [1]{#1}%
\providecommand \bibfnamefont [1]{#1}%
\providecommand \citenamefont [1]{#1}%
\providecommand \href@noop [0]{\@secondoftwo}%
\providecommand \href [0]{\begingroup \@sanitize@url \@href}%
\providecommand \@href[1]{\@@startlink{#1}\@@href}%
\providecommand \@@href[1]{\endgroup#1\@@endlink}%
\providecommand \@sanitize@url [0]{\catcode `\\12\catcode `\$12\catcode
  `\&12\catcode `\#12\catcode `\^12\catcode `\_12\catcode `\%12\relax}%
\providecommand \@@startlink[1]{}%
\providecommand \@@endlink[0]{}%
\providecommand \url  [0]{\begingroup\@sanitize@url \@url }%
\providecommand \@url [1]{\endgroup\@href {#1}{\urlprefix }}%
\providecommand \urlprefix  [0]{URL }%
\providecommand \Eprint [0]{\href }%
\providecommand \doibase [0]{https://doi.org/}%
\providecommand \selectlanguage [0]{\@gobble}%
\providecommand \bibinfo  [0]{\@secondoftwo}%
\providecommand \bibfield  [0]{\@secondoftwo}%
\providecommand \translation [1]{[#1]}%
\providecommand \BibitemOpen [0]{}%
\providecommand \bibitemStop [0]{}%
\providecommand \bibitemNoStop [0]{.\EOS\space}%
\providecommand \EOS [0]{\spacefactor3000\relax}%
\providecommand \BibitemShut  [1]{\csname bibitem#1\endcsname}%
\let\auto@bib@innerbib\@empty
\bibitem [{\citenamefont {Demidov}\ \emph {et~al.}(2018)\citenamefont
  {Demidov}, \citenamefont {Gorbunov},\ and\ \citenamefont
  {Kirpichnikov}}]{Demidov:2017lzf}%
  \BibitemOpen
  \bibfield  {author} {\bibinfo {author} {\bibfnamefont {S.~V.}\ \bibnamefont
  {Demidov}}, \bibinfo {author} {\bibfnamefont {D.~S.}\ \bibnamefont
  {Gorbunov}},\ and\ \bibinfo {author} {\bibfnamefont {D.~V.}\ \bibnamefont
  {Kirpichnikov}},\ }\bibfield  {title} {\bibinfo {title} {{Gravitational waves
  from phase transition in split NMSSM}},\ }\href
  {https://doi.org/10.1016/j.physletb.2018.02.007} {\bibfield  {journal}
  {\bibinfo  {journal} {Phys. Lett. B}\ }\textbf {\bibinfo {volume} {779}},\
  \bibinfo {pages} {191} (\bibinfo {year} {2018})},\ \Eprint
  {https://arxiv.org/abs/1712.00087} {arXiv:1712.00087 [hep-ph]} \BibitemShut
  {NoStop}%
\bibitem [{\citenamefont {Chala}\ \emph {et~al.}(2018)\citenamefont {Chala},
  \citenamefont {Krause},\ and\ \citenamefont {Nardini}}]{Chala:2018ari}%
  \BibitemOpen
  \bibfield  {author} {\bibinfo {author} {\bibfnamefont {M.}~\bibnamefont
  {Chala}}, \bibinfo {author} {\bibfnamefont {C.}~\bibnamefont {Krause}},\ and\
  \bibinfo {author} {\bibfnamefont {G.}~\bibnamefont {Nardini}},\ }\bibfield
  {title} {\bibinfo {title} {{Signals of the electroweak phase transition at
  colliders and gravitational wave observatories}},\ }\href
  {https://doi.org/10.1007/JHEP07(2018)062} {\bibfield  {journal} {\bibinfo
  {journal} {JHEP}\ }\textbf {\bibinfo {volume} {07}},\ \bibinfo {pages}
  {062}},\ \Eprint {https://arxiv.org/abs/1802.02168} {arXiv:1802.02168
  [hep-ph]} \BibitemShut {NoStop}%
\bibitem [{\citenamefont {Ahriche}\ \emph {et~al.}(2019)\citenamefont
  {Ahriche}, \citenamefont {Hashino}, \citenamefont {Kanemura},\ and\
  \citenamefont {Nasri}}]{Ahriche:2018rao}%
  \BibitemOpen
  \bibfield  {author} {\bibinfo {author} {\bibfnamefont {A.}~\bibnamefont
  {Ahriche}}, \bibinfo {author} {\bibfnamefont {K.}~\bibnamefont {Hashino}},
  \bibinfo {author} {\bibfnamefont {S.}~\bibnamefont {Kanemura}},\ and\
  \bibinfo {author} {\bibfnamefont {S.}~\bibnamefont {Nasri}},\ }\bibfield
  {title} {\bibinfo {title} {{Gravitational Waves from Phase Transitions in
  Models with Charged Singlets}},\ }\href
  {https://doi.org/10.1016/j.physletb.2018.12.013} {\bibfield  {journal}
  {\bibinfo  {journal} {Phys. Lett. B}\ }\textbf {\bibinfo {volume} {789}},\
  \bibinfo {pages} {119} (\bibinfo {year} {2019})},\ \Eprint
  {https://arxiv.org/abs/1809.09883} {arXiv:1809.09883 [hep-ph]} \BibitemShut
  {NoStop}%
\bibitem [{\citenamefont {Caprini}\ \emph {et~al.}(2020)\citenamefont {Caprini}
  \emph {et~al.}}]{Caprini:2019egz}%
  \BibitemOpen
  \bibfield  {author} {\bibinfo {author} {\bibfnamefont {C.}~\bibnamefont
  {Caprini}} \emph {et~al.},\ }\bibfield  {title} {\bibinfo {title} {{Detecting
  gravitational waves from cosmological phase transitions with LISA: an
  update}},\ }\href {https://doi.org/10.1088/1475-7516/2020/03/024} {\bibfield
  {journal} {\bibinfo  {journal} {JCAP}\ }\textbf {\bibinfo {volume} {03}},\
  \bibinfo {pages} {024}},\ \Eprint {https://arxiv.org/abs/1910.13125}
  {arXiv:1910.13125 [astro-ph.CO]} \BibitemShut {NoStop}%
\bibitem [{\citenamefont {Caldwell}\ \emph {et~al.}(2022)\citenamefont
  {Caldwell} \emph {et~al.}}]{Caldwell:2022qsj}%
  \BibitemOpen
  \bibfield  {author} {\bibinfo {author} {\bibfnamefont {R.}~\bibnamefont
  {Caldwell}} \emph {et~al.},\ }\bibfield  {title} {\bibinfo {title}
  {{Detection of Early-Universe Gravitational Wave Signatures and Fundamental
  Physics}},\ }\href@noop {} {\  (\bibinfo {year} {2022})},\ \Eprint
  {https://arxiv.org/abs/2203.07972} {arXiv:2203.07972 [gr-qc]} \BibitemShut
  {NoStop}%
\bibitem [{\citenamefont {Moore}\ and\ \citenamefont
  {Prokopec}(1995)}]{Moore:1995ua}%
  \BibitemOpen
  \bibfield  {author} {\bibinfo {author} {\bibfnamefont {G.~D.}\ \bibnamefont
  {Moore}}\ and\ \bibinfo {author} {\bibfnamefont {T.}~\bibnamefont
  {Prokopec}},\ }\bibfield  {title} {\bibinfo {title} {{Bubble wall velocity in
  a first order electroweak phase transition}},\ }\href
  {https://doi.org/10.1103/PhysRevLett.75.777} {\bibfield  {journal} {\bibinfo
  {journal} {Phys. Rev. Lett.}\ }\textbf {\bibinfo {volume} {75}},\ \bibinfo
  {pages} {777} (\bibinfo {year} {1995})},\ \Eprint
  {https://arxiv.org/abs/hep-ph/9503296} {arXiv:hep-ph/9503296} \BibitemShut
  {NoStop}%
\bibitem [{\citenamefont {Espinosa}\ \emph {et~al.}(2010)\citenamefont
  {Espinosa}, \citenamefont {Konstandin}, \citenamefont {No},\ and\
  \citenamefont {Servant}}]{Espinosa:2010hh}%
  \BibitemOpen
  \bibfield  {author} {\bibinfo {author} {\bibfnamefont {J.~R.}\ \bibnamefont
  {Espinosa}}, \bibinfo {author} {\bibfnamefont {T.}~\bibnamefont
  {Konstandin}}, \bibinfo {author} {\bibfnamefont {J.~M.}\ \bibnamefont {No}},\
  and\ \bibinfo {author} {\bibfnamefont {G.}~\bibnamefont {Servant}},\
  }\bibfield  {title} {\bibinfo {title} {{Energy Budget of Cosmological
  First-order Phase Transitions}},\ }\href
  {https://doi.org/10.1088/1475-7516/2010/06/028} {\bibfield  {journal}
  {\bibinfo  {journal} {JCAP}\ }\textbf {\bibinfo {volume} {06}},\ \bibinfo
  {pages} {028}},\ \Eprint {https://arxiv.org/abs/1004.4187} {arXiv:1004.4187
  [hep-ph]} \BibitemShut {NoStop}%
\bibitem [{\citenamefont {Hindmarsh}\ \emph {et~al.}(2021)\citenamefont
  {Hindmarsh}, \citenamefont {L\"uben}, \citenamefont {Lumma},\ and\
  \citenamefont {Pauly}}]{Hindmarsh:2020hop}%
  \BibitemOpen
  \bibfield  {author} {\bibinfo {author} {\bibfnamefont {M.~B.}\ \bibnamefont
  {Hindmarsh}}, \bibinfo {author} {\bibfnamefont {M.}~\bibnamefont {L\"uben}},
  \bibinfo {author} {\bibfnamefont {J.}~\bibnamefont {Lumma}},\ and\ \bibinfo
  {author} {\bibfnamefont {M.}~\bibnamefont {Pauly}},\ }\bibfield  {title}
  {\bibinfo {title} {{Phase transitions in the early universe}},\ }\href
  {https://doi.org/10.21468/SciPostPhysLectNotes.24} {\bibfield  {journal}
  {\bibinfo  {journal} {SciPost Phys. Lect. Notes}\ }\textbf {\bibinfo {volume}
  {24}},\ \bibinfo {pages} {1} (\bibinfo {year} {2021})},\ \Eprint
  {https://arxiv.org/abs/2008.09136} {arXiv:2008.09136 [astro-ph.CO]}
  \BibitemShut {NoStop}%
\bibitem [{\citenamefont {De~Curtis}\ \emph {et~al.}(2022)\citenamefont
  {De~Curtis}, \citenamefont {Rose}, \citenamefont {Guiggiani}, \citenamefont
  {Muyor},\ and\ \citenamefont {Panico}}]{DeCurtis:2022hlx}%
  \BibitemOpen
  \bibfield  {author} {\bibinfo {author} {\bibfnamefont {S.}~\bibnamefont
  {De~Curtis}}, \bibinfo {author} {\bibfnamefont {L.~D.}\ \bibnamefont {Rose}},
  \bibinfo {author} {\bibfnamefont {A.}~\bibnamefont {Guiggiani}}, \bibinfo
  {author} {\bibfnamefont {A.~G.}\ \bibnamefont {Muyor}},\ and\ \bibinfo
  {author} {\bibfnamefont {G.}~\bibnamefont {Panico}},\ }\bibfield  {title}
  {\bibinfo {title} {{Bubble wall dynamics at the electroweak phase
  transition}},\ }\href {https://doi.org/10.1007/JHEP03(2022)163} {\bibfield
  {journal} {\bibinfo  {journal} {JHEP}\ }\textbf {\bibinfo {volume} {03}},\
  \bibinfo {pages} {163}},\ \Eprint {https://arxiv.org/abs/2201.08220}
  {arXiv:2201.08220 [hep-ph]} \BibitemShut {NoStop}%
\bibitem [{\citenamefont {Laurent}\ and\ \citenamefont
  {Cline}(2022)}]{Laurent:2022jrs}%
  \BibitemOpen
  \bibfield  {author} {\bibinfo {author} {\bibfnamefont {B.}~\bibnamefont
  {Laurent}}\ and\ \bibinfo {author} {\bibfnamefont {J.~M.}\ \bibnamefont
  {Cline}},\ }\bibfield  {title} {\bibinfo {title} {{First principles
  determination of bubble wall velocity}},\ }\href@noop {} {\  (\bibinfo {year}
  {2022})},\ \Eprint {https://arxiv.org/abs/2204.13120} {arXiv:2204.13120
  [hep-ph]} \BibitemShut {NoStop}%
\bibitem [{\citenamefont {Bea}\ \emph {et~al.}(2021)\citenamefont {Bea},
  \citenamefont {Casalderrey-Solana}, \citenamefont {Giannakopoulos},
  \citenamefont {Mateos}, \citenamefont {Sanchez-Garitaonandia},\ and\
  \citenamefont {Zilh\~ao}}]{Bea:2021zsu}%
  \BibitemOpen
  \bibfield  {author} {\bibinfo {author} {\bibfnamefont {Y.}~\bibnamefont
  {Bea}}, \bibinfo {author} {\bibfnamefont {J.}~\bibnamefont
  {Casalderrey-Solana}}, \bibinfo {author} {\bibfnamefont {T.}~\bibnamefont
  {Giannakopoulos}}, \bibinfo {author} {\bibfnamefont {D.}~\bibnamefont
  {Mateos}}, \bibinfo {author} {\bibfnamefont {M.}~\bibnamefont
  {Sanchez-Garitaonandia}},\ and\ \bibinfo {author} {\bibfnamefont
  {M.}~\bibnamefont {Zilh\~ao}},\ }\bibfield  {title} {\bibinfo {title}
  {{Bubble wall velocity from holography}},\ }\href
  {https://doi.org/10.1103/PhysRevD.104.L121903} {\bibfield  {journal}
  {\bibinfo  {journal} {Phys. Rev. D}\ }\textbf {\bibinfo {volume} {104}},\
  \bibinfo {pages} {L121903} (\bibinfo {year} {2021})},\ \Eprint
  {https://arxiv.org/abs/2104.05708} {arXiv:2104.05708 [hep-th]} \BibitemShut
  {NoStop}%
\bibitem [{\citenamefont {Itzhaki}\ \emph {et~al.}(1998)\citenamefont
  {Itzhaki}, \citenamefont {Maldacena}, \citenamefont {Sonnenschein},\ and\
  \citenamefont {Yankielowicz}}]{Itzhaki:1998dd}%
  \BibitemOpen
  \bibfield  {author} {\bibinfo {author} {\bibfnamefont {N.}~\bibnamefont
  {Itzhaki}}, \bibinfo {author} {\bibfnamefont {J.~M.}\ \bibnamefont
  {Maldacena}}, \bibinfo {author} {\bibfnamefont {J.}~\bibnamefont
  {Sonnenschein}},\ and\ \bibinfo {author} {\bibfnamefont {S.}~\bibnamefont
  {Yankielowicz}},\ }\bibfield  {title} {\bibinfo {title} {{Supergravity and
  the large N limit of theories with sixteen supercharges}},\ }\href
  {https://doi.org/10.1103/PhysRevD.58.046004} {\bibfield  {journal} {\bibinfo
  {journal} {Phys. Rev. D}\ }\textbf {\bibinfo {volume} {58}},\ \bibinfo
  {pages} {046004} (\bibinfo {year} {1998})},\ \Eprint
  {https://arxiv.org/abs/hep-th/9802042} {arXiv:hep-th/9802042} \BibitemShut
  {NoStop}%
\bibitem [{\citenamefont {Brandhuber}\ \emph {et~al.}(1998)\citenamefont
  {Brandhuber}, \citenamefont {Itzhaki}, \citenamefont {Sonnenschein},\ and\
  \citenamefont {Yankielowicz}}]{Brandhuber:1998er}%
  \BibitemOpen
  \bibfield  {author} {\bibinfo {author} {\bibfnamefont {A.}~\bibnamefont
  {Brandhuber}}, \bibinfo {author} {\bibfnamefont {N.}~\bibnamefont {Itzhaki}},
  \bibinfo {author} {\bibfnamefont {J.}~\bibnamefont {Sonnenschein}},\ and\
  \bibinfo {author} {\bibfnamefont {S.}~\bibnamefont {Yankielowicz}},\
  }\bibfield  {title} {\bibinfo {title} {{Wilson loops, confinement, and phase
  transitions in large N gauge theories from supergravity}},\ }\href
  {https://doi.org/10.1088/1126-6708/1998/06/001} {\bibfield  {journal}
  {\bibinfo  {journal} {JHEP}\ }\textbf {\bibinfo {volume} {06}},\ \bibinfo
  {pages} {001}},\ \Eprint {https://arxiv.org/abs/hep-th/9803263}
  {arXiv:hep-th/9803263} \BibitemShut {NoStop}%
\bibitem [{\citenamefont {Witten}(1998)}]{Witten:1998zw}%
  \BibitemOpen
  \bibfield  {author} {\bibinfo {author} {\bibfnamefont {E.}~\bibnamefont
  {Witten}},\ }\bibfield  {title} {\bibinfo {title} {{Anti-de Sitter space,
  thermal phase transition, and confinement in gauge theories}},\ }\href
  {https://doi.org/10.4310/ATMP.1998.v2.n3.a3} {\bibfield  {journal} {\bibinfo
  {journal} {Adv. Theor. Math. Phys.}\ }\textbf {\bibinfo {volume} {2}},\
  \bibinfo {pages} {505} (\bibinfo {year} {1998})},\ \Eprint
  {https://arxiv.org/abs/hep-th/9803131} {arXiv:hep-th/9803131} \BibitemShut
  {NoStop}%
\bibitem [{\citenamefont {Horowitz}\ and\ \citenamefont
  {Myers}(1998)}]{Horowitz:1998ha}%
  \BibitemOpen
  \bibfield  {author} {\bibinfo {author} {\bibfnamefont {G.~T.}\ \bibnamefont
  {Horowitz}}\ and\ \bibinfo {author} {\bibfnamefont {R.~C.}\ \bibnamefont
  {Myers}},\ }\bibfield  {title} {\bibinfo {title} {{The AdS / CFT
  correspondence and a new positive energy conjecture for general
  relativity}},\ }\href {https://doi.org/10.1103/PhysRevD.59.026005} {\bibfield
   {journal} {\bibinfo  {journal} {Phys. Rev. D}\ }\textbf {\bibinfo {volume}
  {59}},\ \bibinfo {pages} {026005} (\bibinfo {year} {1998})},\ \Eprint
  {https://arxiv.org/abs/hep-th/9808079} {arXiv:hep-th/9808079} \BibitemShut
  {NoStop}%
\bibitem [{\citenamefont {Aharony}\ \emph {et~al.}(2007)\citenamefont
  {Aharony}, \citenamefont {Sonnenschein},\ and\ \citenamefont
  {Yankielowicz}}]{Aharony:2006da}%
  \BibitemOpen
  \bibfield  {author} {\bibinfo {author} {\bibfnamefont {O.}~\bibnamefont
  {Aharony}}, \bibinfo {author} {\bibfnamefont {J.}~\bibnamefont
  {Sonnenschein}},\ and\ \bibinfo {author} {\bibfnamefont {S.}~\bibnamefont
  {Yankielowicz}},\ }\bibfield  {title} {\bibinfo {title} {{A Holographic model
  of deconfinement and chiral symmetry restoration}},\ }\href
  {https://doi.org/10.1016/j.aop.2006.11.002} {\bibfield  {journal} {\bibinfo
  {journal} {Annals Phys.}\ }\textbf {\bibinfo {volume} {322}},\ \bibinfo
  {pages} {1420} (\bibinfo {year} {2007})},\ \Eprint
  {https://arxiv.org/abs/hep-th/0604161} {arXiv:hep-th/0604161} \BibitemShut
  {NoStop}%
\bibitem [{\citenamefont {Aharony}\ \emph {et~al.}(2006)\citenamefont
  {Aharony}, \citenamefont {Minwalla},\ and\ \citenamefont
  {Wiseman}}]{Aharony:2005bm}%
  \BibitemOpen
  \bibfield  {author} {\bibinfo {author} {\bibfnamefont {O.}~\bibnamefont
  {Aharony}}, \bibinfo {author} {\bibfnamefont {S.}~\bibnamefont {Minwalla}},\
  and\ \bibinfo {author} {\bibfnamefont {T.}~\bibnamefont {Wiseman}},\
  }\bibfield  {title} {\bibinfo {title} {{Plasma-balls in large N gauge
  theories and localized black holes}},\ }\href
  {https://doi.org/10.1088/0264-9381/23/7/001} {\bibfield  {journal} {\bibinfo
  {journal} {Class. Quant. Grav.}\ }\textbf {\bibinfo {volume} {23}},\ \bibinfo
  {pages} {2171} (\bibinfo {year} {2006})},\ \Eprint
  {https://arxiv.org/abs/hep-th/0507219} {arXiv:hep-th/0507219} \BibitemShut
  {NoStop}%
\bibitem [{\citenamefont {Janik}\ \emph {et~al.}(2021)\citenamefont {Janik},
  \citenamefont {J{\"a}rvinen},\ and\ \citenamefont
  {Sonnenschein}}]{Janik:2021jbq}%
  \BibitemOpen
  \bibfield  {author} {\bibinfo {author} {\bibfnamefont {R.~A.}\ \bibnamefont
  {Janik}}, \bibinfo {author} {\bibfnamefont {M.}~\bibnamefont
  {J{\"a}rvinen}},\ and\ \bibinfo {author} {\bibfnamefont {J.}~\bibnamefont
  {Sonnenschein}},\ }\bibfield  {title} {\bibinfo {title} {{A simple
  description of holographic domain walls in confining theories \textemdash{}
  extended hydrodynamics}},\ }\href {https://doi.org/10.1007/JHEP09(2021)129}
  {\bibfield  {journal} {\bibinfo  {journal} {JHEP}\ }\textbf {\bibinfo
  {volume} {09}},\ \bibinfo {pages} {129}},\ \Eprint
  {https://arxiv.org/abs/2106.02642} {arXiv:2106.02642 [hep-th]} \BibitemShut
  {NoStop}%
\bibitem [{\citenamefont {Bigazzi}\ \emph {et~al.}(2020)\citenamefont
  {Bigazzi}, \citenamefont {Caddeo}, \citenamefont {Cotrone},\ and\
  \citenamefont {Paredes}}]{Bigazzi:2020phm}%
  \BibitemOpen
  \bibfield  {author} {\bibinfo {author} {\bibfnamefont {F.}~\bibnamefont
  {Bigazzi}}, \bibinfo {author} {\bibfnamefont {A.}~\bibnamefont {Caddeo}},
  \bibinfo {author} {\bibfnamefont {A.~L.}\ \bibnamefont {Cotrone}},\ and\
  \bibinfo {author} {\bibfnamefont {A.}~\bibnamefont {Paredes}},\ }\bibfield
  {title} {\bibinfo {title} {{Fate of false vacua in holographic first-order
  phase transitions}},\ }\href {https://doi.org/10.1007/JHEP12(2020)200}
  {\bibfield  {journal} {\bibinfo  {journal} {JHEP}\ }\textbf {\bibinfo
  {volume} {12}},\ \bibinfo {pages} {200}},\ \Eprint
  {https://arxiv.org/abs/2008.02579} {arXiv:2008.02579 [hep-th]} \BibitemShut
  {NoStop}%
\bibitem [{\citenamefont {Ares}\ \emph
  {et~al.}(2022{\natexlab{a}})\citenamefont {Ares}, \citenamefont {Henriksson},
  \citenamefont {Hindmarsh}, \citenamefont {Hoyos},\ and\ \citenamefont
  {Jokela}}]{Ares:2021ntv}%
  \BibitemOpen
  \bibfield  {author} {\bibinfo {author} {\bibfnamefont {F.~R.}\ \bibnamefont
  {Ares}}, \bibinfo {author} {\bibfnamefont {O.}~\bibnamefont {Henriksson}},
  \bibinfo {author} {\bibfnamefont {M.}~\bibnamefont {Hindmarsh}}, \bibinfo
  {author} {\bibfnamefont {C.}~\bibnamefont {Hoyos}},\ and\ \bibinfo {author}
  {\bibfnamefont {N.}~\bibnamefont {Jokela}},\ }\bibfield  {title} {\bibinfo
  {title} {{Effective actions and bubble nucleation from holography}},\ }\href
  {https://doi.org/10.1103/PhysRevD.105.066020} {\bibfield  {journal} {\bibinfo
   {journal} {Phys. Rev. D}\ }\textbf {\bibinfo {volume} {105}},\ \bibinfo
  {pages} {066020} (\bibinfo {year} {2022}{\natexlab{a}})},\ \Eprint
  {https://arxiv.org/abs/2109.13784} {arXiv:2109.13784 [hep-th]} \BibitemShut
  {NoStop}%
\bibitem [{\citenamefont {Ares}\ \emph
  {et~al.}(2022{\natexlab{b}})\citenamefont {Ares}, \citenamefont {Henriksson},
  \citenamefont {Hindmarsh}, \citenamefont {Hoyos},\ and\ \citenamefont
  {Jokela}}]{Ares:2021nap}%
  \BibitemOpen
  \bibfield  {author} {\bibinfo {author} {\bibfnamefont {F.~R.}\ \bibnamefont
  {Ares}}, \bibinfo {author} {\bibfnamefont {O.}~\bibnamefont {Henriksson}},
  \bibinfo {author} {\bibfnamefont {M.}~\bibnamefont {Hindmarsh}}, \bibinfo
  {author} {\bibfnamefont {C.}~\bibnamefont {Hoyos}},\ and\ \bibinfo {author}
  {\bibfnamefont {N.}~\bibnamefont {Jokela}},\ }\bibfield  {title} {\bibinfo
  {title} {{Gravitational Waves at Strong Coupling from an Effective Action}},\
  }\href {https://doi.org/10.1103/PhysRevLett.128.131101} {\bibfield  {journal}
  {\bibinfo  {journal} {Phys. Rev. Lett.}\ }\textbf {\bibinfo {volume} {128}},\
  \bibinfo {pages} {131101} (\bibinfo {year} {2022}{\natexlab{b}})},\ \Eprint
  {https://arxiv.org/abs/2110.14442} {arXiv:2110.14442 [hep-th]} \BibitemShut
  {NoStop}%
\bibitem [{\citenamefont {Bantilan}\ \emph {et~al.}(2020)\citenamefont
  {Bantilan}, \citenamefont {Figueras},\ and\ \citenamefont
  {Mateos}}]{Bantilan:2020pay}%
  \BibitemOpen
  \bibfield  {author} {\bibinfo {author} {\bibfnamefont {H.}~\bibnamefont
  {Bantilan}}, \bibinfo {author} {\bibfnamefont {P.}~\bibnamefont {Figueras}},\
  and\ \bibinfo {author} {\bibfnamefont {D.}~\bibnamefont {Mateos}},\
  }\bibfield  {title} {\bibinfo {title} {{Real-time Dynamics of Plasma Balls
  from Holography}},\ }\href {https://doi.org/10.1103/PhysRevLett.124.191601}
  {\bibfield  {journal} {\bibinfo  {journal} {Phys. Rev. Lett.}\ }\textbf
  {\bibinfo {volume} {124}},\ \bibinfo {pages} {191601} (\bibinfo {year}
  {2020})},\ \Eprint {https://arxiv.org/abs/2001.05476} {arXiv:2001.05476
  [hep-th]} \BibitemShut {NoStop}%
\bibitem [{\citenamefont {Janik}\ \emph {et~al.}(2017)\citenamefont {Janik},
  \citenamefont {Jankowski},\ and\ \citenamefont
  {Soltanpanahi}}]{Janik:2017ykj}%
  \BibitemOpen
  \bibfield  {author} {\bibinfo {author} {\bibfnamefont {R.~A.}\ \bibnamefont
  {Janik}}, \bibinfo {author} {\bibfnamefont {J.}~\bibnamefont {Jankowski}},\
  and\ \bibinfo {author} {\bibfnamefont {H.}~\bibnamefont {Soltanpanahi}},\
  }\bibfield  {title} {\bibinfo {title} {{Real-Time dynamics and phase
  separation in a holographic first order phase transition}},\ }\href
  {https://doi.org/10.1103/PhysRevLett.119.261601} {\bibfield  {journal}
  {\bibinfo  {journal} {Phys. Rev. Lett.}\ }\textbf {\bibinfo {volume} {119}},\
  \bibinfo {pages} {261601} (\bibinfo {year} {2017})},\ \Eprint
  {https://arxiv.org/abs/1704.05387} {arXiv:1704.05387 [hep-th]} \BibitemShut
  {NoStop}%
\bibitem [{\citenamefont {Bellantuono}\ \emph {et~al.}(2019)\citenamefont
  {Bellantuono}, \citenamefont {Janik}, \citenamefont {Jankowski},\ and\
  \citenamefont {Soltanpanahi}}]{Bellantuono:2019wbn}%
  \BibitemOpen
  \bibfield  {author} {\bibinfo {author} {\bibfnamefont {L.}~\bibnamefont
  {Bellantuono}}, \bibinfo {author} {\bibfnamefont {R.~A.}\ \bibnamefont
  {Janik}}, \bibinfo {author} {\bibfnamefont {J.}~\bibnamefont {Jankowski}},\
  and\ \bibinfo {author} {\bibfnamefont {H.}~\bibnamefont {Soltanpanahi}},\
  }\bibfield  {title} {\bibinfo {title} {{Dynamics near a first order phase
  transition}},\ }\href {https://doi.org/10.1007/JHEP10(2019)146} {\bibfield
  {journal} {\bibinfo  {journal} {JHEP}\ }\textbf {\bibinfo {volume} {10}},\
  \bibinfo {pages} {146}},\ \Eprint {https://arxiv.org/abs/1906.00061}
  {arXiv:1906.00061 [hep-th]} \BibitemShut {NoStop}%
\bibitem [{Note1()}]{Note1}%
  \BibitemOpen
  \bibinfo {note} {The models differ in the intermediate region where the
  Witten's model shows wavy behavior. Apparently the waves are absent in the
  nonconformal model because it includes dissipation, whereas for the Witten's
  model we use a perfect fluid description.}\BibitemShut {Stop}%
\bibitem [{\citenamefont {Landau}\ and\ \citenamefont
  {Lifshitz}(2013)}]{landau2013fluid}%
  \BibitemOpen
  \bibfield  {author} {\bibinfo {author} {\bibfnamefont {L.}~\bibnamefont
  {Landau}}\ and\ \bibinfo {author} {\bibfnamefont {E.}~\bibnamefont
  {Lifshitz}},\ }\href {https://books.google.es/books?id=CeBbAwAAQBAJ} {\emph
  {\bibinfo {title} {Fluid Mechanics}}},\ \bibinfo {series} {Course of
  Theoretical Physics}\ No.\ \bibinfo {number} {v. 6}\ (\bibinfo  {publisher}
  {Elsevier Science},\ \bibinfo {year} {2013})\BibitemShut {NoStop}%
\bibitem [{\citenamefont {Gyulassy}\ \emph {et~al.}(1984)\citenamefont
  {Gyulassy}, \citenamefont {Kajantie}, \citenamefont {Kurki-Suonio},\ and\
  \citenamefont {McLerran}}]{Gyulassy:1983rq}%
  \BibitemOpen
  \bibfield  {author} {\bibinfo {author} {\bibfnamefont {M.}~\bibnamefont
  {Gyulassy}}, \bibinfo {author} {\bibfnamefont {K.}~\bibnamefont {Kajantie}},
  \bibinfo {author} {\bibfnamefont {H.}~\bibnamefont {Kurki-Suonio}},\ and\
  \bibinfo {author} {\bibfnamefont {L.~D.}\ \bibnamefont {McLerran}},\
  }\bibfield  {title} {\bibinfo {title} {{Deflagrations and Detonations as a
  Mechanism of Hadron Bubble Growth in Supercooled Quark Gluon Plasma}},\
  }\href {https://doi.org/10.1016/0550-3213(84)90004-X} {\bibfield  {journal}
  {\bibinfo  {journal} {Nucl. Phys. B}\ }\textbf {\bibinfo {volume} {237}},\
  \bibinfo {pages} {477} (\bibinfo {year} {1984})}\BibitemShut {NoStop}%
\bibitem [{Note2()}]{Note2}%
  \BibitemOpen
  \bibinfo {note} {The pressure of the perfect fluid wave is close to the
  pressure inside the bubble, hence its temperature is typically higher and
  closer to $T_c$ than the temperature of the asymptotic overcooled
  plasma.}\BibitemShut {Stop}%
\bibitem [{\citenamefont {Bea}\ \emph {et~al.}(2022)\citenamefont {Bea},
  \citenamefont {Casalderrey-Solana}, \citenamefont {Giannakopoulos},
  \citenamefont {Jansen}, \citenamefont {Mateos}, \citenamefont
  {Sanchez-Garitaonandia},\ and\ \citenamefont {Zilh\~ao}}]{Bea:2022mfb}%
  \BibitemOpen
  \bibfield  {author} {\bibinfo {author} {\bibfnamefont {Y.}~\bibnamefont
  {Bea}}, \bibinfo {author} {\bibfnamefont {J.}~\bibnamefont
  {Casalderrey-Solana}}, \bibinfo {author} {\bibfnamefont {T.}~\bibnamefont
  {Giannakopoulos}}, \bibinfo {author} {\bibfnamefont {A.}~\bibnamefont
  {Jansen}}, \bibinfo {author} {\bibfnamefont {D.}~\bibnamefont {Mateos}},
  \bibinfo {author} {\bibfnamefont {M.}~\bibnamefont {Sanchez-Garitaonandia}},\
  and\ \bibinfo {author} {\bibfnamefont {M.}~\bibnamefont {Zilh\~ao}},\
  }\bibfield  {title} {\bibinfo {title} {{Holographic Bubbles with Jecco:
  Expanding, Collapsing and Critical}},\ }\href@noop {} {\  (\bibinfo {year}
  {2022})},\ \Eprint {https://arxiv.org/abs/2202.10503} {arXiv:2202.10503
  [hep-th]} \BibitemShut {NoStop}%
\bibitem [{\citenamefont {Bigazzi}\ \emph {et~al.}(2021)\citenamefont
  {Bigazzi}, \citenamefont {Caddeo}, \citenamefont {Canneti},\ and\
  \citenamefont {Cotrone}}]{Bigazzi:2021ucw}%
  \BibitemOpen
  \bibfield  {author} {\bibinfo {author} {\bibfnamefont {F.}~\bibnamefont
  {Bigazzi}}, \bibinfo {author} {\bibfnamefont {A.}~\bibnamefont {Caddeo}},
  \bibinfo {author} {\bibfnamefont {T.}~\bibnamefont {Canneti}},\ and\ \bibinfo
  {author} {\bibfnamefont {A.~L.}\ \bibnamefont {Cotrone}},\ }\bibfield
  {title} {\bibinfo {title} {{Bubble wall velocity at strong coupling}},\
  }\href {https://doi.org/10.1007/JHEP08(2021)090} {\bibfield  {journal}
  {\bibinfo  {journal} {JHEP}\ }\textbf {\bibinfo {volume} {08}},\ \bibinfo
  {pages} {090}},\ \Eprint {https://arxiv.org/abs/2104.12817} {arXiv:2104.12817
  [hep-ph]} \BibitemShut {NoStop}%
\end{thebibliography}%

\clearpage
\appendix

\section{Supplemental material}

\subsection{Details on the simplified model}

We discuss here some details of the hydrodynamic model of~\eqref{e.simp1}. In this Lagrangian we have chosen the units such that the critical temperature of the deconfinement transition $T_c=1$, and the pressure, which is that of a conformal theory, is given by $p(T)=T^4$. Natural choices for the coefficients $c$ and $d$ are power laws in temperature with powers depending in $\gm$. 
In~\cite{Janik:2021jbq} we argued that the choice
\eq \label{e.cdcoeffs}
 c(T,\gm) = c\, T^{\hat \al(1+\Gamma(\gm))} \ , \ \  d(T,\gm) = \frac{1}{2}c\,q_*^2\, T^{2(1-\hat \al)(1+\Gamma(\gm))}
\eqx
for these coefficients gives a precise fit to the numerical AMW domain wall solution. Here the coefficient $\hat \al$ can take any value. The constants $c$ and $q_*$, which control the value of the surface tension and the width of the domain wall respectively, were fitted to the AMW solution. This fit gives $q_*/(2\pi T_c) \approx 0.682$, but because the value of $q_*$ can be absorbed into rescalings of the space-time coordinates, we can set $q_*=1$ without loss of generality. For this choice, the AMW value for $c$ is $c \approx 5.892$. For the numerical solutions in the Witten model in this letter, we also set $\hat \al = 0$. Only fine details of the solutions are affected by this choice.
Notice that  the dependence on $\gm$ is absent at the critical temperature:
\eq
 c(1,\gm) = c \ , \qquad d(1,\gm) = \frac{1}{2}c\,q_*^2
\eqx
so in this sense the dependence on $\gm$ of the coefficients is weak.

The expressions for the energy-momentum tensors of the confined and deconfined phases in~\eqref{e.simp3} are given as
\eqn
 T_{\mu\nu}^\mathrm{confining} &=& \eta_{\mu\nu} = \mathrm{diag}(-1,1,1) \\
 T_{\mu\nu}^\mathrm{deconfined} &=& p(T)\eta_{\mu\nu} + 4 p(T) u_\mu u_\nu
\eqnx
where $u_\mu$ is the fluid velocity satisfying $\eta^{\mu\nu}u_\mu u_\nu=-1$ and we eliminated the energy density by using the relation $\eps(T) =3 p(T)$ for conformal theories. Here we chose to write the tensor for the 2+1 dimensional system, i.e., we did not write down the components in the third, compactified spatial direction. The space-time coordinates are then $(t,x,y)$. We will choose the planar domain wall to lie at fixed value of the coordinate $x$, so that our solutions will be independent of $y$. At critical temperature and for fluid at rest the deconfined tensor $T_{\mu\nu}^\mathrm{deconfined} = \mathrm{diag}(3,1,1)$ in our conventions. For the ``domain wall'' contribution to the energy-momentum tensor we find
\eqn
 T_{\mu\nu}^\Sg &=& c(T,\gm) \partial_\mu\gm\partial_\nu\gm \\\nonumber
 &&-\frac{1}{2} \left[c(T,\gm)(\partial \gm)^2+ d(T,\gm)\gm^2(1-\gm)^2\right]\eta_{\mu\nu} \\\nonumber
 && -T\left[(\partial \gm)^2\partial_T c(T,\gm) +\gm^2(1-\gm)^2\partial_Td(T,\gm)\right] u_\mu u_\nu\\
 &=& c\, \partial_\mu\gm\partial_\nu\gm \\\nonumber
 &&-\frac{c}{2} \left[(\partial \gm)^2+ q_*^2\,T^{2(1+\Gamma(\gm))}\,\gm^2(1-\gm)^2\right]\eta_{\mu\nu} \\\nonumber
 && -c\, q_*^2\, \left(1+\Gamma(\gm)\right)\,T^{2(1+\Gamma(\gm))}\,\gm^2(1-\gm)^2 u_\mu u_\nu
\eqnx
where for the second expression we assumed~\eqref{e.cdcoeffs} with $\hat \al=0$.

\subsection{Hydrodynamic simple wave}

For completeness we will present here the derivation of the simple wave formula (\ref{e.nonlinear}) adapted from \cite{landau2013fluid}.
The simple wave is defined by the requirement that all hydrodynamic variables are expressed in terms of one of them, e.g. the pressure $p$, which in turn depends on the $t$ and $x$ coordinates.
Hence we have
\eq
T^{\mu\nu} = (\eps(p)+p)u^\mu u^\nu +p \eta^{\mu\nu} \quad u^\mu=(\cosh \al(p), \sinh\al(p), 0)
\eqx
The energy-momentum conservation equations become
\eq
\partial_\mu p \, \partial_p T^{\mu\nu}=0
\eqx
The existence of a nontrivial solution requires that the determinant of the matrix $\partial_p T^{\mu\nu}$ vanishes. This yields
\eq
\partial_p T^{tt} \partial_p T^{xx} - \left(\partial_p T^{tx}\right)^2 =0
\eqx
which reduces to
\eq
\partial_p \al(p) = \f{\sqrt{\eps'(p)}}{\eps(p) + p} \equiv \f{1}{(\eps + p)c_s}
\eqx
and gives directly the integral formula (using $c_s^2=\f{1}{\eps'(p)}$).

\end{document}